# Long-term Solar Activity Studies using Microwave Imaging Observations and Prediction for Cycle 25

N. Gopalswamy[1*], P. Mäkelä[1,2], S. Yashiro[1,2], S. Akiyama[1,2]

[1]Code 671, NASA Goddard Space Flight Center, Greenbelt, Maryland, USA

[2]Department of Physics, The Catholic University of America, Washington DC 20064

*Corresponding author, e-mail: nat.gopalswamy@nasa.gov

**Abstract**

We use microwave imaging observations from the Nobeyama Radioheliograph at 17 GHz for long-term studies of solar activity. In particular, we use the polar and low-latitude brightness temperatures as proxies to the polar magnetic field and the active-regions, respectively. We also use the location of prominence eruptions as a proxy to the filament locations as a function of time. We show that the polar microwave brightness temperature is highly correlated with the polar magnetic field strength and the fast solar wind speed. We also show that the polar microwave brightness at one cycle is correlated with the low latitude brightness with a lag of about half a solar cycle. We use this correlation to predict the strength of the solar cycle: the smoothed sunspot numbers in the southern and northern hemispheres can be predicted as 89 and 59, respectively. These values indicate that cycle 25 will not be too different from cycle 24 in its strength. We also combined the rush to the pole data from Nobeyama prominences with historical data going back to 1860 to study the north-south asymmetry of sign reversal at solar poles. We find that the reversal asymmetry has a quasi-periodicity of 3-5 cycles.

## 1. Introduction

Understanding the long-term variability is a key aspect of assessing the anticipated space weather impact. The level of solar activity not only determines the frequency and intensity of transient events, but also the severity of their consequences. A prominent example is the mild space weather in solar cycle (SC) 24 owing to the weak solar activity as indicated by the reduced sunspot number (SSN). The weak solar activity also significantly altered the physical conditions of the heliosphere (McComas et al. 2013; Gopalswamy et al. 2014), which in turn significantly affected the physical properties of coronal mass ejections (CMEs) and hence their space weather impact (Gopalswamy et al. 2014; 2015). The idea of forecasting solar cycles has a long history (Ohl 1968), but the use of polar field strength to predict the strength of the following cycle was first advocated by Schatten et al. (1978). This suggestion is based on the Babcock-Leighton mechanism: the magnetic flux from the trailing regions of sunspots reaching the pole, reversing it, and building it with the new polarity that maximizes during the minimum phase of a solar cycle; the polar flux sinks to the bottom of the convection zone, and the resulting poloidal field is then stretched by the differential rotation and emerges as the toroidal field (sunspots) (see Charbonneau 2010; Petrie, 2015). Schatten et al. (1978) used four different methods of estimating the polar field strength and found a linear correlation between the inferred polar field strength of one cycle and the maximum SSN of the following cycle. One of these methods uses the number of polar faculae compiled by Sheeley (1964). Makarov and Makarova (1996) reported that the monthly number of polar faculae correlates with the monthly sunspot area with a time shift of about 6 years. The extended and refined compilation of polar faculae (Sheeley 2008) has been shown to be a reliable proxy of the polar field strength (Munoz-Jaramillo et al. 2013). The close connection between polar faculae and the polar field strength has been recently demonstrated using Hinode observations: magnetic patches in the polar region host faculae that possess kG magnetic fields (Kaithakkal et al. 2013).

In this paper, we present another signature of the polar magnetic field strength – the polar brightness temperature (Tb) enhancement above the quiet Sun level using the Nobeyama radioheliograph (NoRH) images at 17 GHz (Nakajima et al. 1994). The microwave Tb enhancement was found to be associated with the magnetic field patches at the bottom of coronal holes (Kosugi et al. 1986). The microwave Tb was found to be correlated with the strength of the magnetic field both in the equatorial (Gopalswamy et al. 2000) and polar (Gopalswamy et al. 2012) coronal holes. The microwave brightness enhancement (MBE) is likely associated with the polar faculae (Koshiishi 1996; Selhorst et al. 2003) because of the enhanced magnetic field in them. It must be noted that the MBE more closely corresponds to the enhanced magnetic patches inside coronal holes rather than the entire coronal holes (Gopalswamy et al. 1999; 2000; 2012; Akiyama et al. 2013). Polar Tb peaks during solar minima and attain their minimum value (quiet Sun Tb ~$10^4$ K) during solar maxima. Thus the polar Tb follows an 11-year cycle, which is 180° out of phase with the sunspot cycle, similar to the number of polar faculae (Makarov and Makarova 1996). Based on a study showing that the MBE of low-latitude coronal holes is correlated with the magnetic field strength in the coronal holes (Gopalswamy et al. 2000) it was found that the polar MBE is a sensitive indicator of the polar field strength (Gopalswamy et al. 2012). In this paper, we revisit the polar MBE and use it to estimate the strength of solar cycle 25.

The growth of the polar fields build is closely related to the latitude distribution of filaments and hence is important in understanding the long-term behavior of solar magnetism. In particular, the high-latitude (HL) filaments that form the polar crown are of interest because they occur only during the maximum phase of solar cycles. Filaments migrate from mid latitudes to the polar region during the maximum phase in a process described as Rush to the Poles (RTTP), which is known from late 1800s (see historical references in Ananthakrishnan 1952; Stix 1974). After the discovery that the sign reversal at solar poles occurs in the maximum phase of the solar cycle (Babcock 1959; 1961), it became evident that the reversal coincides with the end of RTTP (Waldmeier 1960; Hyder 1965). The reversal was also found to be asymmetric: the south pole reversed first and then the north pole in solar cycle 19. To use as a proxy to the epoch of polarity reversal relevant to dynamo models, Stix (1974) had compiled RTTP data from various sources for solar cycles 10-20. We make use of this compilation to study the reversal asymmetry by extending Stix's (1974) compilation using data from Fujimori (1984), Lorenc et al. (2003),

Pojoga and Huang (2003), and Gopalswamy et al. (2003; 2012). The extended data base allows us to study the reversal asymmetry over more than 150 years from 1860 to the present. The reversal asymmetry has been suggested to be a consequence of hemispheric asymmetry of sunspot activity (Svalgaard and Kamide 2013), but the recent reversal in cycle 24 has shown that the sign reversal is more complicated.

**2. Data**

In this study, we use two sets of data: (i) the polar microwave Tb obtained from NoRH images at 17 GHz, and (ii) the locations of prominence eruptions (PEs) detected automatically from the NoRH images (from 1992 onwards) and from the 304 Å images obtained by the Atmospheric Imaging Assembly (AIA, Lemen et al. 2012) on board the Solar Dynamics Observatory (SDO) from 2010 onwards. Figure 1 shows a composite plot of the two data sets: a super synoptic chart known as microwave butterfly diagram (Gopalswamy et al. 2012) with superposed PE latitudes. The microwave butterfly diagram essentially shows the latitudinal distribution of microwave Tb as a function of time. The basic data used in the construction of the butterfly diagram are the daily-best microwave images taken around local noon (3 UT) at 17 GHz. These images have a spatial resolution of about 10" and distinguish features such as dark filaments on the disk, bright prominences on the limb, bright active regions, and bright polar cap when there is a coronal hole present. The daily images are used to build Carrington rotation maps, which are then averaged over Carrington rotation (CR) periods and assembled in a time sequence to get the butterfly diagram.

The microwave butterfly diagram shows three pairs of low-latitude (LL) patches that correspond to the active region belt similar to sunspot or magnetic butterfly diagrams (Hathaway, 2010). The patches correspond to three solar cycles: part of cycle 22 (leftmost pair), whole of cycle 23 (middle pair), and cycle 24 (rightmost pair). The NoRH observations started in the declining phase of cycle 22 (July 1992) and hence the incomplete LL brightness patches. The middle LL pair (cycle 23) shows the familiar drift of the active regions toward the equator. The Tb contours of cycle 24 (right LL pair) indicate that that the cycle is much weaker than cycle 23. During the maximum phase, cycle 23 shows multiple episodes of high activity. In cycle 24, there is only one episode in each hemisphere corresponding to the two peaks in SSN (2012 in the north and 2014 in the south). The polar MBE, unique to microwave frequencies in the range 15-90 GHz is related to the enhanced magnetic field in coronal holes (Gopalswamy et al. 1999a; 2000). Although the physical mechanism of the enhancement is not fully understood (see e.g., Gopalswamy et al. 1999a,b; Nindos et al. 1999; Gopalswamy et al. 2000; 2012; Selhorst et al. 2003; Prosovetsky and Myagkova, 2011; Shibasaki (2013), it is a definite indicator of enhanced magnetic field strength in coronal holes. We shall investigate how the polar MBE is related to (i) the observed polar magnetic field strength, (ii) the solar wind speed from the polar regions, (iii) the peak SSN in the next cycle, and the peak low-latitude Tb in the next cycle. Based on the level of polar Tb in cycle 24, we shall predict the strength of cycle 25.

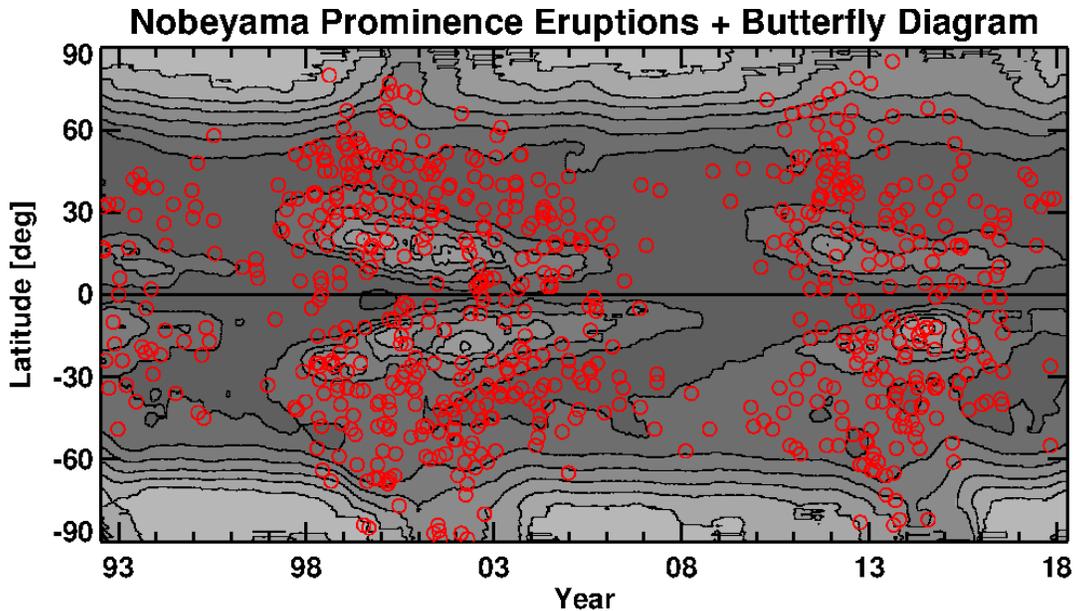

Figure 1. Locations (red circles) of NoRH microwave prominence eruptions (PEs) superposed on the most recent microwave butterfly diagram (contours and grayscale). A 13-rotation smoothing has been applied to the NoRH brightness temperature (Tb) along the time axis to eliminate the periodic variation due to solar B0-angle variation. The contour levels are at 10,000, 10,300, 10,609, 10,927, 11,255, 11,592, and 11,940 K. The HL MBE patches correspond to times of enhanced magnetic field strength during solar minima. The LL MBE patches correspond to active regions that become prominent during solar maxima. Thus the microwave butterfly diagrams resemble the magnetic butterfly diagram, except that the sign of the magnetic field is not in the microwave butterfly diagram (updated from Gopalswamy et al. 2012).

The superposed PE locations in Fig. 1 indicate that they are generally situated between the LL and HL Tb enhancements (i.e., mid latitudes). The PEs in the active region belt are part of low-latitude CMEs that originate in sunspot regions. When active regions emerge closer to the equator in the declining phase, the PE locations also do so. During the maximum phases of solar cycles, the PE locations go beyond the 60º latitude, representing the RTTP phenomenon related to streams of magnetic flux moving toward the poles from the trailing parts of active regions. These PEs are due to the eruption filaments formed at the edges of the flux surges where they meet opposite polarity regions. Gopalswamy et al. (2003) showed that PEs can be used as proxies for filaments in tracking the poleward movement of filament locations. This is a way to track the RTTP without observing all the filaments. We shall combine the NoRH PE data with the historical RTTP data to study the north-south asymmetry in the reversal.

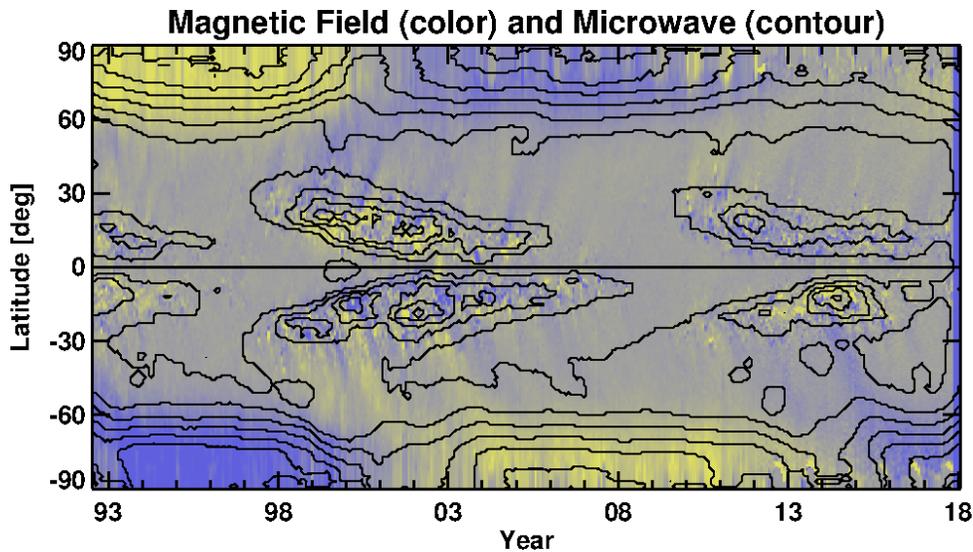

Figure 2. Microwave butterfly diagram (contours) overlaid on the SOLIS magnetic butterfly diagram since the beginning of NoRH operation in July 1992. Yellow and blue colors denote positive and negative polarities, respectively. A 13-rotation smoothing has been used along the time axis to eliminate the periodic variation due to solar B0-angle variation. The contour levels are at 10,000, 10,300, 10,609, 10,927, 11,255, 11,592, and 11,940 K. Flux surges from active regions can be seen towards the poles. Note that the polar microwave brightness patches correspond to unipolar magnetic field regions.

We also use SOLIS (Synoptic Optical Long-term Investigations of the Sun) magnetograms obtained by the National Solar Observatory under the Integrated Synoptic Program (NISP). These data are needed to show the close correspondence between microwave Tb and the magnetic field strength at the photospheric level. Throughout this paper, we use SSN (Version 2.0) data (both total and hemispheric) from SILSO (Sunspot Index and Long-term Solar Observations), Royal Observatory of Belgium. In comparing microwave Tb to the solar wind speed, we use data from the Solar Wind Over the Poles of the Sun (SWOOPS) instrument on board the Ulysses spacecraft (McComas et al. 2000). Gaps in the SOLIS data have been filled using magnetograms from SOHO's Michelson Doppler Imager (MDI, Scherrer et al., 1995) and SDO's Helioseismic and Magnetic Imager (HMI, Scherrer et al., 2012). In order to illustrate the utility of polar microwave Tb for precursor-based SSN prediction, we use the aa index from the National Geophysical Data Center (NGDC, ftp://ftp.ngdc.noaa.gov/STP/GEOMAGNETIC_DATA/AASTAR/aaindex) until the end of 2010. For the period after 2010, we us the aa index available at the British Geological Survey (http://www.geomag.bgs.ac.uk/data_service/data/home.html).

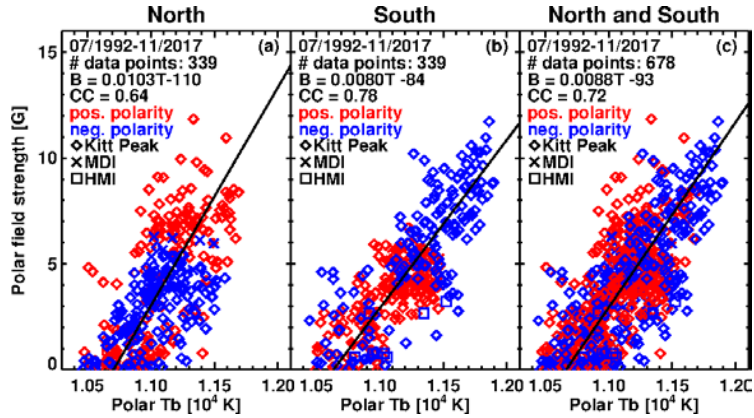

Figure 3. Scatter plots between the polar microwave Tb and the polar magnetic field strength averaged over latitudes poleward of $60°$ plotted for the northern (a) and southern (b) and both (c) hemispheres. The data are from 339 CRs from 1992 until November 2017 (each data point corresponds to one CR). There were two intervals of peak polar B and Tb in each hemisphere as indicated by the colors. Only the magnitudes of B are plotted, but the signs (red positive and blue negative) are indicated by the colors. Gaps in the SOLIS data were filled using data from SOHO/MDI and SDO/HMI.

## 3 Analysis of Polar Microwave Brightness Temperature Distribution

### 3.1 Polar Microwave Tb and Field Strength

In order to show the correspondence between polar microwave Tb and the polar field strength, we have shown a composite butterfly diagram using SOLIS magnetic data (color) and NoRH Tb (contours) in Fig. 2. We see that the polar patches of enhanced microwave Tb correspond remarkably to the unipolar magnetic regions that switch the polarity every 11 years around solar maxima. We can also see poleward surges of magnetic flux transported from the trailing portions of active regions. These surges typically take about 2 years to reach the poles. Figure 3 shows scatter plots between the magnetic field strength B and Tb, averaged over latitudes poleward of $60°$, separately for the northern and southern hemispheres. Each data point corresponds to the average values in one CR. Different colors (red – positive; blue – negative) in the same hemisphere correspond to different polar patches (see Fig. 2). In the southern hemisphere we have data points from three different patches, while in two patches in the northern hemisphere (the third patch in the north polar region is not significantly developed). The correlation coefficients are 0.64 (northern hemisphere) and 0.78 (southern hemisphere) for 339 data points in each hemisphere. The correlation is highly significant because the Pearson critical correlation coefficient (the probability that the observed correlation is by chance) is only 0.179 for $p=5\times10^{-4}$. The combined data set has a correlation coefficient of 0.72 for 678 data points, which is highly significant (Pearson critical correlation coefficient is 0.132; $p = 5\times10^{-4}$). The regression line for the combined data set is $B = 0.0088Tb – 93$ G, where Tb is in K. The slope is slightly larger than the one for the period July 1992 to March 2012: $B = 0.0067Tb –72$ G (Gopalswamy et al. 2012). The difference is mainly due to the fitting program used in the earlier work, which considered errors only in the independent variable. The fitting used here accounts for errors in both X and Y variables (Isobe et al. 1990).

## 3.2 Relation between Polar Microwave Tb and Solar Wind Speed

The close correlation between the polar B at the photosperic level and the microwave Tb suggests that there must be a relation between the latter and the solar wind speed. This is because the the solar wind speed Vsw from coronal holes is known to be correlated with the underlying phtospheric field strength (Kojima et al. 2004; Fisk et al. 1999). A better correlation is known between Vsw and B/f where f is the flux tube expansion factor from the photosphere to the source surface (Fujiki et al. 2005; Akiyama et al. 2013), but here we consider only Tb, which is a proxy to B as we demonstrated above. In order to check the Tb – Vsw correlation, we make use of the fortunate cirumstances that all the Ulysses high-latitude passes overlapped with NoRH observations. The Ulyssess polar passes in the first and third orbits were during solar minima: cycle 22/23 minimum (south: June to November 1994 and north: June to September 1995) and cycle 23/24 minimum (south: November 2006 to April 2007 and north: November 2007 to March 2008). The second orbit had polar passes during the cycle 23 maximum (south: September 2000 to January 2001 and north: September to December 2001).

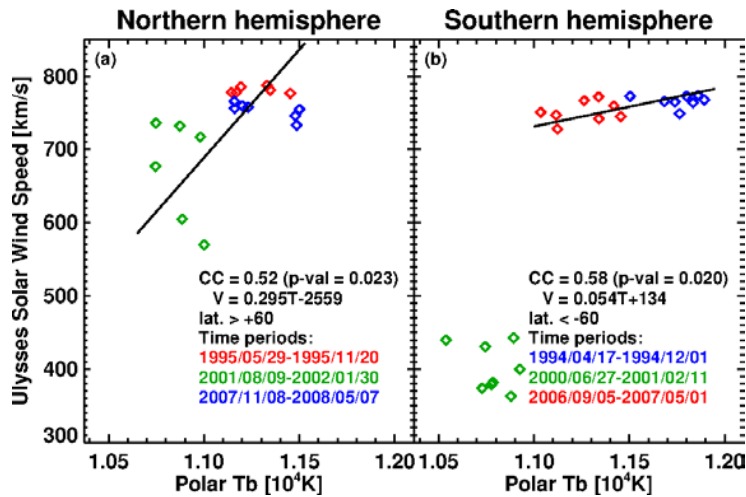

Figure 4. Correlation between high-latitude brightness temperature at the Sun and the solar wind speed measured by Ulysses at high latitudes for three different epochs when Ulysses was observing from latitudes poleward of 60º for the northern (a) and southern (b) hemispheres. Both the brightness temperature Tb and Vsw at Ulysses are averaged over CR periods. The number of rotations in the correlation: 19 in the north and 16 in the south. The correlation coefficients are significant, well above the Pearson's critical correlation coefficient. The p-values are noted on the plots. Fast wind takes about 10 days to reach Ulysses. We did not take this into account in the correlation because the Carrington rotation period is longer than the wind travel time.

In Fig. 4 we have plotted the solar wind speed measured by the Ulysses/SWOOPS instrument averaged over CR periods as a function of the polar microwave Tb. The measurements correspond to the three passes when the Ulysses spacecraft was poleward of 60º latitude. High speed wind was observed during all the three northern polar passes, while only during two of the southern polar passes (during the two solar minima). The south pole of the Sun had maximum conditions during the second pass (the maximum phase lasted until middle of 2002), so no fast wind was observed (The wind speed was only around 400 km/s). There was also no south polar

Tb enhancement (see Fig. 2) at this time. On the other hand, the maximum condition had already ended in the north pole by September 2000 and the new polarity magnetic field had started building up during the north polar pass as also seen in the Tb enhnacement. That means the fast wind had already started blowing above the north pole of the Sun.  There is a moderate correlation between Vsw and Tb both in the northern and southern hemispheres, considering only the fast wind. The correlations are 0.52 and 0.58 in the northern and southern  hemispheres, respectively. These are significant because the correlation coefficients exceed the corresponding Pearson critical coefficients (0.456 for 19 data points in the north and 0.497 for 16 data points in the south) and the p-values are also very low.  Thus the polar microwave Tb is a definite indicator of fast solar wind speed, consistent with the observations (Kojima et al. 2004) and theory (Fisk et al. 1999).

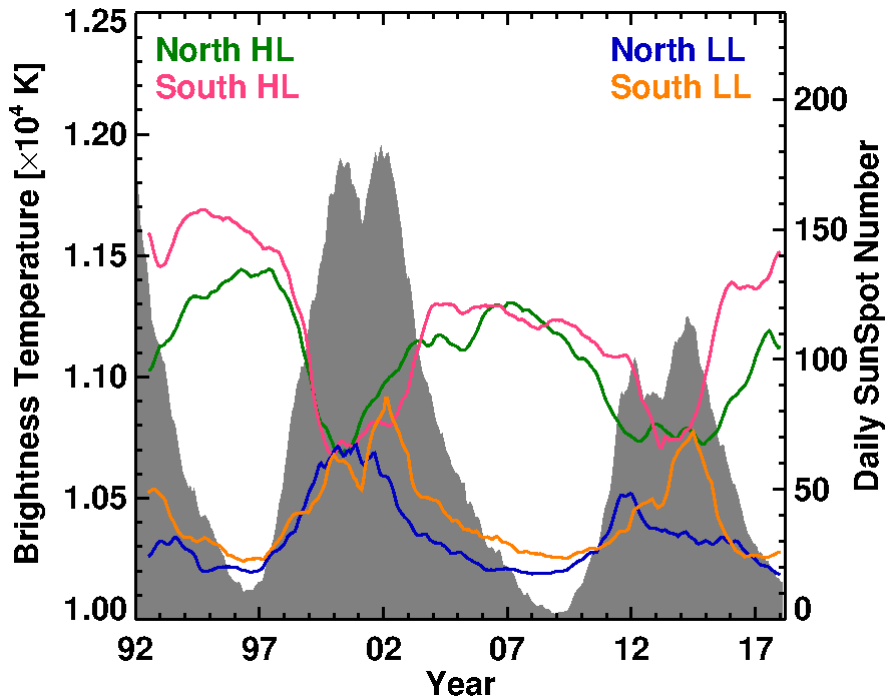

Figure 5. Polar (HL) and low-latitude (LL) microwave Tb compared with the sunspot number (gray). The HL Tb is averaged over latitudes poleward of 60º. The LL Tb is averaged over the latitude range 0-40º. This selection of latitude ranges avoids any overlap between LL and HL Tb values. The Tb values are plotted separately for northern and southern hemispheres. The anticorrelation between HL and LL Tb values is obvious.

### 3.3 Polar Microwave Tb and Solar Cycle Prediction

Another implication of the close relation between the polar mirowave Tb enhancement and the polar field strength is that the former should indicate the strength of the next solar cycle, according to the precursor method (see e.g., Petrovay 2010).  The precursor method estimates the polar magnetic field strength using various techniques during the minimum phase of a given cycle and uses it to predict the strength of the next cycle. In order to demonstrate the utility of polar microwave Tb, we have shown plots of the HL and LL Tb from NoRH in Fig. 5 with the SSN shown for reference. The plotted HL Tb values are derived from Fig. 2 by averaging Tb values over latitudes poleward of 60º; the LL Tb values are averaged over the latitude range 0 to 40º. The data are also smoothed over 13 CRs to eliminate the modulation caused by the solar B0 angle. As in Fig. 2, we see a close correspondence beween SSN and LL Tb, including the double peak in SSN. Even though the Tb data correspond to a little more than two solar cycles, we clearly see that LL and HL Tb are anticorrelated in a given cycle, but the HL Tb of one cycle is postively correlated with the LL Tb of the next cycle. This is true for both northern and southern hemispheres. The HL Tb is not sharply peaked at the sunspot minimum, somewhat similar to the flat axial dipole moment reported by Iijima et al. (2017).

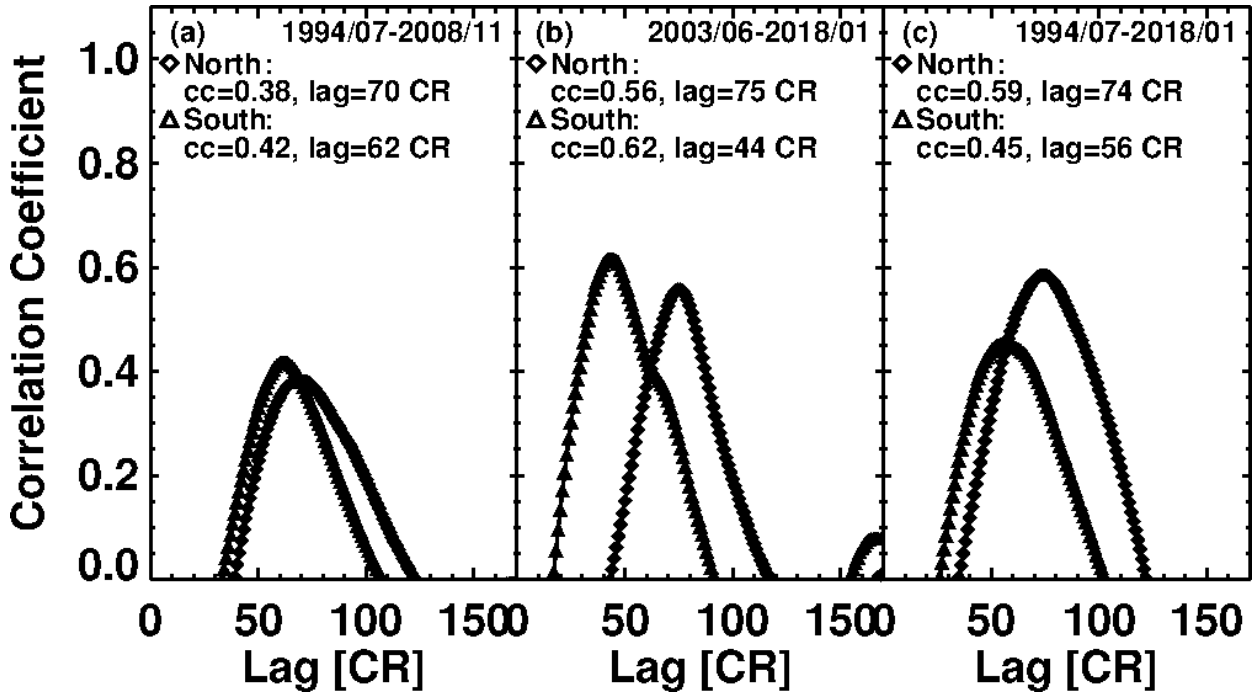

Figure 6. Results of cross-correlation analysis between HL and LL microwave Tb. (a) for the period July 1994 to November 2008 showing the correlation between the HL Tb from 22/23 minimum and LL Tb in cycle 23, (b) June 2003 to January 2018 showing the correlation between HL Tb from 23/24 minimum and LL Tb in cycle 24, and (cc) correlation between HL and LL Tb's for the combined period (July 1994 to January 2018). The peak correlation coefficients and the lags at the maximum correlation are noted on the plots.

In order to show the correlation between the HL and LL Tb, we performed a cross correlation analysis using the available LL and HL Tb data. The peak correlation is obtained when the lag is 70-75 CRs in the northern hemisphere and 44-62 CRs in the southern hemisphere. These lags roughly corrspond to half a solar cycle. Some correlations are low but statistically significant in all cases because of the large number of data points. One obvious feature in Fig. 6 is that the lag in the northern hemisphere is longer in both cycles. The correlations are consistent with the Babcock-Leighton mechanism of the solar cycle. These results suggest that the peak HL Tb during the minimum phase of a solar cycle can be used to predict the strength of the next cycle, similar to the use of aa index during solar minima (see, e.g., Wang and Sheeley 2009; Svalgaard et al. 2005). The aa index is a three-hourly index of geomagnetic activity determined from the *k* indices scaled at two antipodal subauroral stations: Canberra in Australia, and Hartland in England. During solar maxima, CMEs dominate the geomagnetic variability, while in the declining phase, low-latitude coronal holes are the drivers via corotating interaction regions. During the minimum phase, the variations are caused by Sun's polar field, which bends down close to the equator and sampled by Earth. Therefore the aa index near the minimum phase of solar cycle is a good measure of Sun's peak polar field strength. Figure 7 shows the relation of SSN with the HL Tb and the average aa index. The aa index used in this figure is the daily values averaged over CR periods. Note the lowest value of the aa index during solar minima roughly occurs slighlty delayed (by ~ 1 year) with respect to the SSN minimum and roughly coincides with the peak HL Tb. Unlike the aa index, the polar microwave Tb has only two primary features: the broad peak during solar minimum phases and a narrower drop during solar maximum phases. Furthermore, the polar microwave Tb is sensitive to polar B in individual solar hemispheres (Fig. 3).

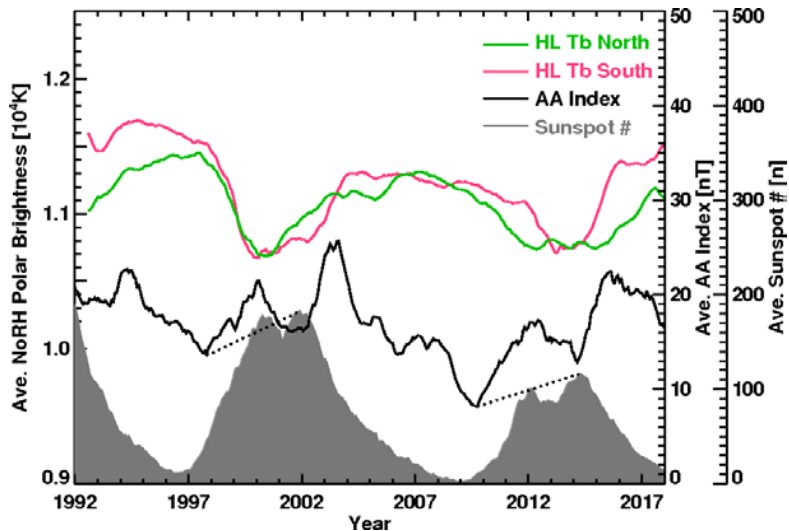

Figure 7. Comparison betweeen the HL microwave Tb and the aa index averaged over CR periods from July 1992 to December 2017. HL Tb is shown separately for the northern and southern hemispheres. The average SSN (daily values averaged over CR periods) is shown in gray for refrence. The dotted line indiates that the aa index at a solar minimum is related to the maximum SSN of the next cycle.

The correlation between the HL Tb during cycle 22/23 and 23/24 minima with the peak daily SSN averaged over CR periods is shown in Fig. 8. The northern and southern hemispheres are treated independently. We have used Tb and SSN averaged over Carrington rotations and those smoother over 13 rotations. The correlation is positive and high for both smoothed and unsmoothed cases. We appreciate that the number of data points is too small, so the confidence level is not high. However, the trend is consistent with other studies that relate the strength of a solar cycle with a proxy of the polar field strength in a preceding cycle (e.g., Schatten et al. 1978) in accordance with the Babcock-Leighton mechanism of the solar cycle (Babcock, 1961; Leigton 1964). In the correlation plots reported by Schatten et al. (1978) only two or three data points were used. Recently, Iijima et al. (2017) used only 3 data points in the correlation of the axial dipole moment at the end of a solar cycle to the peak sunspot number in the next solar cycle.

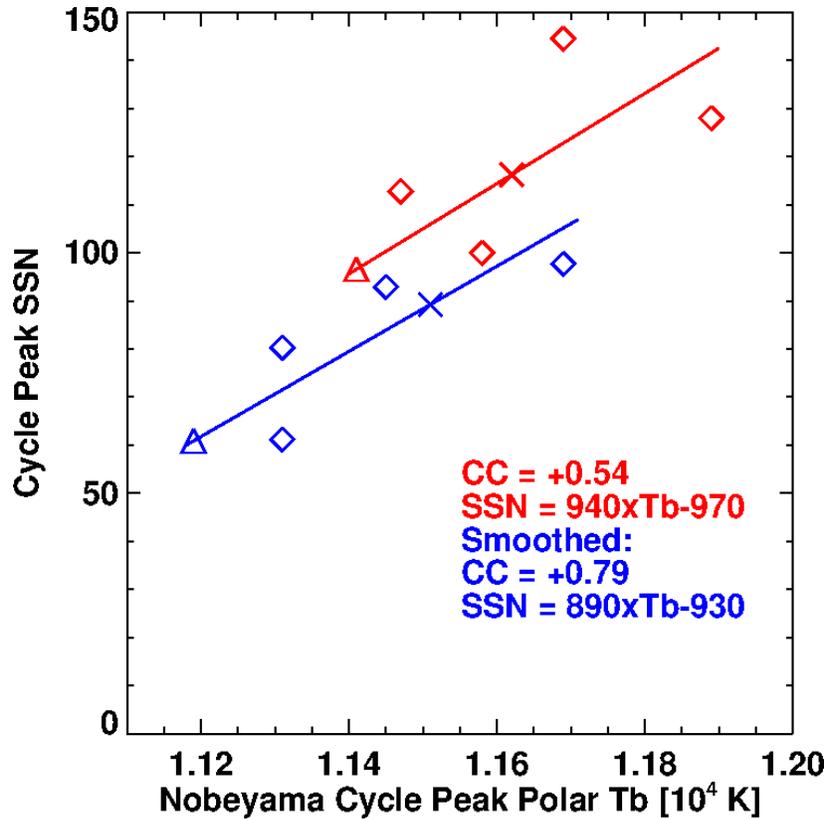

Figure 8. Scatterplot between the peak HL Tb during the minimum phase of a solar cycle n and the peak hemispheric SSN in the next solar cycle (n+1). There were two solar minima during the interval 1992 to 2018 (cycle 22/23 and 23/24). The four data points correspond to the north and south polar brightness temperatures for the two minima. The peak values obtained from unsmoothed (red) and 13-rotation smoothed (blue) data are shown separately. The correlations are high, but the statistical significance is low because there are only four data points. However, the correlation shows the right trend, given the fact that the polar microwave Tb is highly correlated with the polar magnetic field strength as shown in Fig. 3. The plus and triangle symbols on the regression lines are the predicted SSN values for cycle 25 in the southern and northern hemispheres, respectively.

As for cycle 25, we can attempt to predict its strength in the southern hemisphere based on the regression lines in Fig. 8. The south-polar peak Tb is $1.162 \times 10^4$ K attained during CR 2196. This is expected to result in a CR-averaged SSN in cycle 25 to be 116.2 in the southern hemisphere. If we use the regression line for the smoothed data, the peak Tb is $1.151 \times 10^4$ K during CR 2199 that is expected to result in a smoothed sunspot number of 89.2. The corresponding numbers in cycle 24 were 112.8 and 80.3 for unsmoothed and smoothed data, respectively. These results indicate that the strength of cycle 25 is expected to be similar to that of cycle 24 in the southern hemisphere. An important caveat is that these estimates assume that the polar microwave Tb may not increase significantly from the current values. In the northern hemisphere, the current level of Tb is probably not the peak value because the reversal started quite late. The current maximum Tb values are $1.141\times10^4$ K (unsmoothed, in CR 2189) and $1.119\times10^4$ K (smoothed, in CR 2193). From these values, we can estimate lower limits to the expected SSN in the northern hemisphere to be 96.5 and 58.7, respectively. Thus the predicted cycle-25 SSN values are also similar to those observed in cycle 24: 100.0 (unsmoothed) and 61.2 (smoothed). The analysis based on the polar microwave Tb thus predicts that the cycle 25 strength is not too different from that of cycle 24. We note that the prediction by Iijima et al. (2017) for the 13-month smoothed total SSN is slightly below 80. This is similar to our smoothed cases if we average over the two hemispheres (77.6). The similarity is understandable because the microwave Tb does not have a sharp peak, but a plateau as in the case of axial dipole moment in Iijima et al. (2017).

## 4. Analysis of the Rush to the Poles

High-latitude prominences/filaments and their poleward migration were discovered by Secchi in 1872. The poleward migration was also known as "dash to the poles" (Evershed and Evershed 1917) and "rush to the poles" (RTTP, Ananthakrishnan 1954). RTTP refers to filament locations systematically moving toward the poles in both hemispheres (Lockyer 1931). RTTP was graphically demonstrated by Ananthakrishnan (1952) for the period from 1905 to 1950 (for cycles 14–18) and is a consistent feature in every solar cycle (Waldmeier 1973; Stix 1974; Fujimori 1984; Lorenc et al. 2003; McIntosh 2003; Gopalswamy et al. 2003; 2012) – as consistent as the Schwabe cycle (see Cliver 2014 for a review). While most of the previous works dealt with prominences or filaments, Gopalswamy et al. (2003; 2012) introduced the locations of prominence eruptions (PEs) as proxy to the locations of prominences/filaments.

After the discovery of the polarity reversal at solar poles (Babcock 1959), Waldmeier (1960) and Hyder (1965) demonstrated the synchronism between RTTP and the sign reversal at solar poles (see also Howard and Labonte 1981). Babcock (1959) noted that the south pole reversed first during March to July 1957, whereas the north reversed in November 1958. Hyder (1965) indicated the end of RTTP in early 1957 in the south and late 1958 in the north, although there was a second branch in the north that indicated an RTTP end in late 1959. In fact our detailed examination of the daily prominence/filament drawings from the Kodaikanal Solar Observatory indicated that the RTTP ended in August 1959 in the south and October 1959 in the north. Despite the 1-2 year uncertainty in the exact time of reversal, these observations indicated that neither the reversals nor the ends of RTTP were in phase in the two hemispheres. This north-south asymmetry provides important clues to the understanding of the variability in solar activity, especially over decadal time scales.

Stix (1974) compiled RTTP data from cycles 10-20 to show that the end of RTTP can be considered as the time of polarity reversal. However, Stix (1974) argued that the polarity reversal need not be in the maximum phase of a cycle based on the report of Gillespie et al (1973) that the south pole in cycle 20 reversed only in the middle of the year 1972. He also pointed out that by 1972, the sunspot activity declined significantly, approaching the next minimum. A detailed examination of the Kodaikanal prominence data revealed that there was sustained high-latitude prominence activity only until 1970 November 8 in the southern hemisphere. Isolated prominences were observed a few times at southern high latitudes (see Table 1). No prominences were observed in the days immediately before and after the ones listed in Table 1. Such isolated high-latitude prominences are not uncommon: in the plot published by Lorenc et al. (2003), one can see such isolated prominences in most of the cycles. Waldmeier (1973) plotted the prominence areas in the northern and southern high latitudes and found that the southern hemispheric prominence area disappeared towards the end of 1970 (their Fig.1b), consistent with the Kodaikanal data.

Table 1. Isolated high-latitude prominences in the southern hemisphere of cycle 20 after the end of RTTP in November 1970. The longitude of the prominence is noted (E-East; W-West)

| Date | Latitude Range (longitude) |
|---|---|
| 1970 November 19 | 71-72 (E) |
| 1971 August 15 | 85-86 (E) |
| 1971 November 6 | 71-73 (E) |
| 1972 May 21 | 80-82 (W) |
| 1972 September 23 | 74-75 (W) |
| 1972 October 20 | 73-74 (E) |
| 1972 November 9 | 74-75 (E), 75-76 (W) |

It must be noted that the end of RTTP is a prerequisite for polarity reversal because the presence of high-latitude prominences indicates the presence of bipolar regions (formed by the polarities of incumbent and insurgent magnetic fluxes). End of RTTP signifies the time of zero polar field. The field in the new polarity will take some time to build to a level significant enough to be observed and hence once expects some delay between the end of RTTP and the time the polarity reversal is observed in magnetograms. It must be appreciated that it is difficult to observe the weak polar magnetic field especially because the field is generally perpendicular to the line of sight. During the well observed reversal in cycle 24, Gopalswamy et al. (2016) found that the reversal was complete within 6 months to a year from the time the zero-field condition was attained at the solar poles. The end of high-latitude prominence activity was much closer (within six months) to the time of polarity reversal.

Even though Stix (1974) did not consider the north-south asymmetry in the polarity reversal, his compilation of RTTP tracks for cycles 10-20 is useful for our purposes. Figure 9 shows the RTTP phenomena compiled by Stix (1974) augmented by newer observations from Fujimori (1984), Lorenc et al. (2003), Pojoga and Huang (2003), and Gopalswamy et al. (2003; 2012). The RTTP asymmetry is indicated by the circled pairs of arrow marks in Fig. 9. In the 15 solar cycles plotted, 12 of them showed definite RTTP asymmetry between the northern and southern hemispheres. Prominence tracks were available only in the trailing part of cycle 10 after the RTTP had ended. The ends of RTTP seem to be simultaneous in cycles 11, 13, and 15, so the north-south asymmetry is inconclusive. Fortunately, for cycles 13 and 15, there are other published reports of prominence tracks. Evershed and Evershed (1917) reported on the prominence migration for the period 1890 to 1914 using prominence observations from Kodaikanal Solar Observatory. Their plot (see Fig. 5 in Cliver 2014) shows RTTP tracks for cycles 13 and 14. Examination of this plot revealed that RTTP ended first in the north and a few months later in the south. This is also confirmed by another plot of prominence migrations reported in from Bocchino (1933) during cycles 12–15 (see Fig. 6 in Cliver 2014). For the cycle 15 reversal, the daily prominence area plotted by Ananthakrishnan (1952) shows that RTTP ended in the north 2-3 months into the year 1918. In the south, RTTP ended 7-8 months into 1918, indicating a 5-month delay in the southern hemisphere. Thus, we have definite RTTP (and hence reversal) asymmetry information continuously for cycles 12-24. One of the remarkable results in Fig. 9 is that the reversal asymmetry shows a quasi-periodicity: after every few cycles, the asymmetry switched between north/south (NS) to south/north (SN). During cycles 12-15, the reversal order was NS; during cycles 16-20 it was SN; during cycles 21-23 it was back to NS before switching to SN in cycle 24. The switching occurred during the maxima of cycles 16, 21, and 24 – that is three switches in 14 cycles.

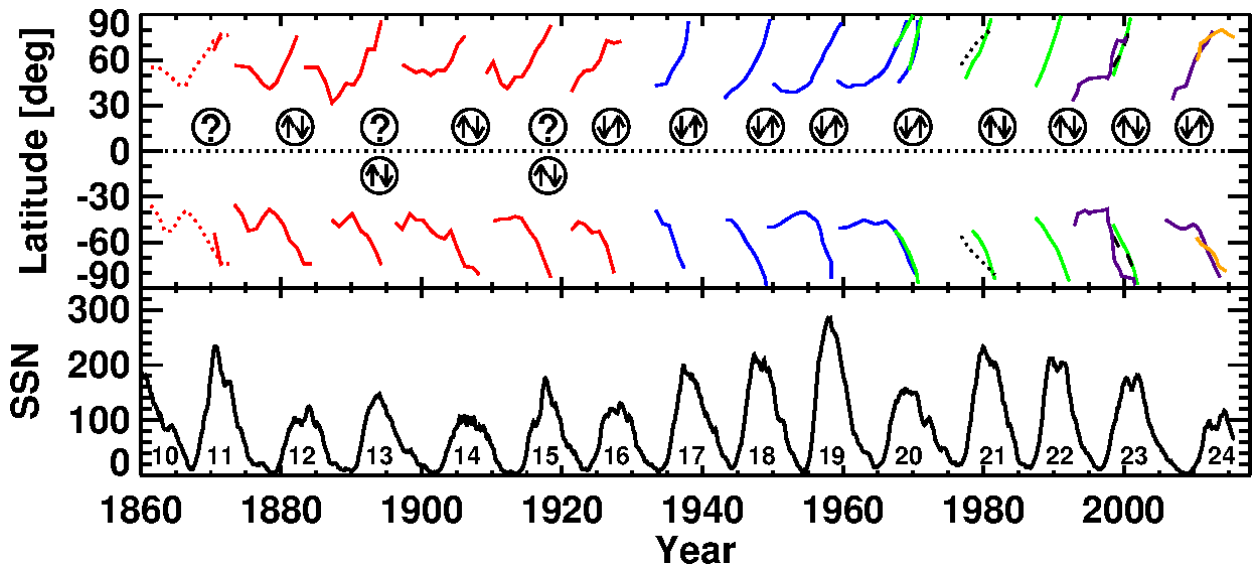

Figure 9. (top) Time-latitude plots of the polar prominences/filaments showing their poleward migration since 1860. Data have been taken from multiple sources. Magenta lines mark the most recent data obtained from locations of prominence eruptions detected automatically from the NoRH images. The orange lines are from PEs detected automatically from SDO/AIA images at 304 Å (see also Fig. 10). Earlier data are taken from published figures: red lines (Lockyer 1931); blue lines (Stix 1974); green lines (Lorenc et al. 2003); dotted black lines (Fujimori 1984); and dashed black lines (Pojoga and Huang 2003). The arrows inside circles indicate at which order the "rush to the pole" ended: northern (arrow up) and southern (arrow down) hemispheres. During the maximum of cycles 13 and 15, the end of RTTP is simultaneous in the north and south, so there was apparently no reversal asymmetry (as indicated by the question mark). However, reports in the published literature (Evershed and Evershed 1917; Ananthakrishnan 1952) clearly showed that the RTTP activity ended first in the north and then in the south as indicated by the circled arrow marks directly below the question marks. (bottom) Version 2.0 of the 13-month smoothed monthly total SSN provided by WDC-SILSO, Royal Observatory of Belgium, Brussels for solar cycles 10 through 24 (as noted).

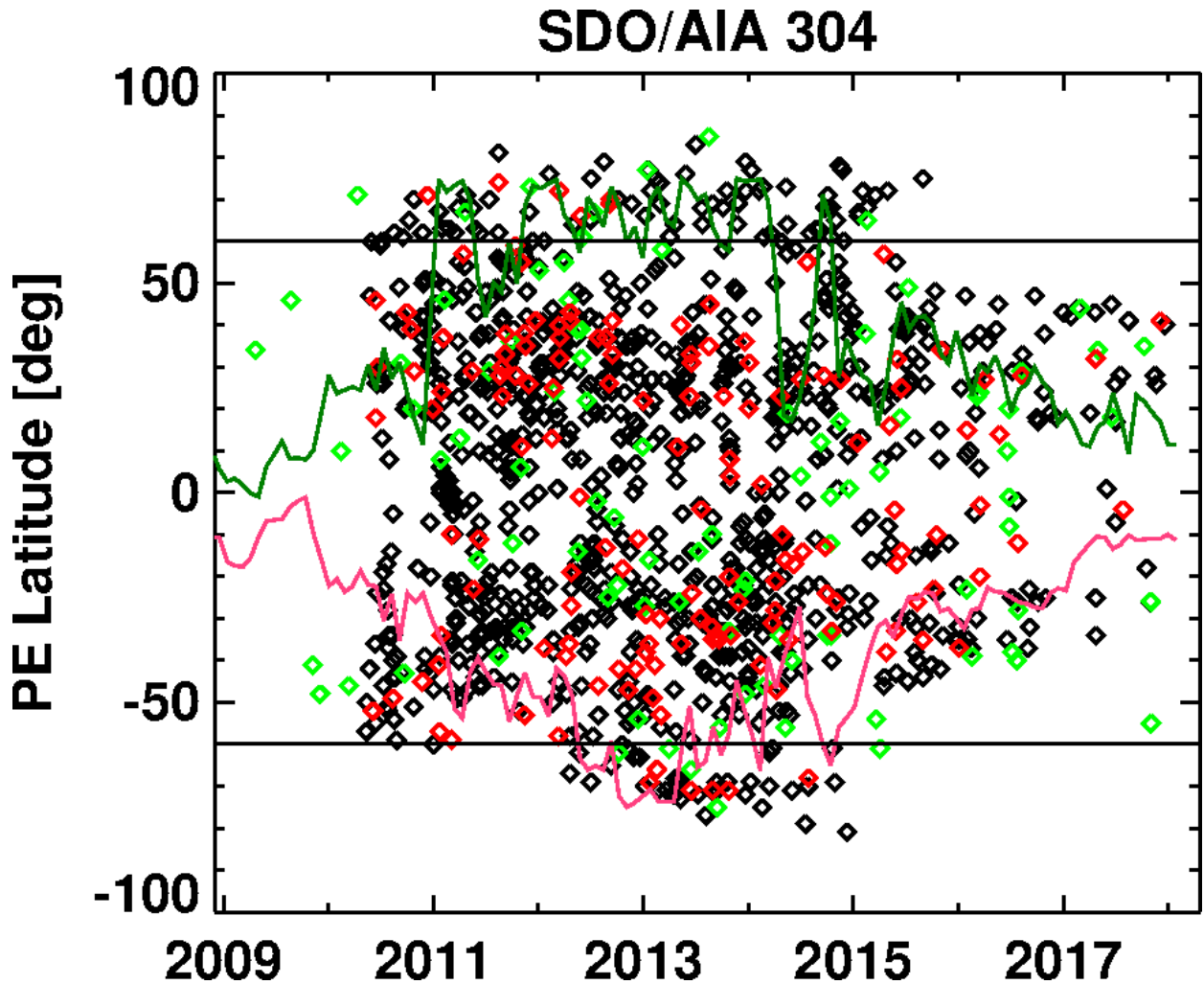

Figure 10. Time-latitude plot of the PE locations detected automatically in NoRH 17 GHz and SDO/AIA 304 Å images. Black: PEs detected exclusively by SDO, red: PEs detected by both SDO and NoRH, and green: PEs detected only by NoRH. The tilt angle of the heliospheric current sheet is overlaid on the plot (solid curves). The horizontal lines mark ±60º latitudes. The last northern PE event was observed by SDO/AIA between 16:38 and 17:00 UT on 2015 August 30 from N75E05. This PE occurred outside the NoRH observing window. The last PE from the south occurred on 2014 December 11 at 01:09 UT. The PE was back sided, located at S81.

Svalgaard and Kamide (2013) investigated the reversal asymmetry in cycles 21-23. In these cycles the reversal asymmetry was NS. Unfortunately, the new pattern emerged only in cycle 21 (the reversal pattern was SN type until cycle 20). They anticipated a NS asymmetry in cycle 24 although they did not have information over the full cycle. However, it turned out that the reversal in the north was not complete until the middle of 2015 because the high-latitude PE activity did not cease until that time. Even though the RTTP activity started a bit late in the southern hemisphere, it ended quickly towards the end of 2014 due to a strong surge associated with the second peak in the sunspot number. In the northern hemisphere, the prolonged high-latitude PE activity was sustained by surges of alternating polarities that thwarted the reversal until the end of 2015 (Gopalswamy et al. 2016).

Figure 10 shows the presence of sustained high-latitude PEs obtained from NoRH and SDO/AIA 304 Å images. The tilt angle of the heliospheric current sheet superposed on the PE location plot also shows the extended period of high tilt angles. The cessation of high-latitude PE activity was on 2015 August 30 in the northern hemisphere and 2014 December 11 in the southern hemisphere, signaling the SN-type reversal. Interestingly, the reversal asymmetry also switched in cycle 16, during which there was prolonged presence of high-latitude prominences (see Fig. 9) as in cycle 24. The prolonged zero-field condition before reversal has been shown to be due to surges that violate Joy's law (Sun et al. 2015; Mordvinov et al. 2016; Gopalswamy et al. 2016). Such surges cause the reversal to be episodic instead of being sharp. In an ideal situation, where all the emerging active regions follow Joy's law, one expects the polar reversal asymmetry to follow the sunspot hemispheric asymmetry as suggested by Svalgaard and Kamide (2013). In reality, there is a large scatter in the distribution of tilt angles that introduce large variability in the polar field strength attained during the minimum phase, thus causing the variability in the amplitude and phase of the new cycle (Dasi-Espuig et al. 2010; Jiang et al. 2014; Cameron and Schüssler 2017; Nagy et al. 2017). In other words, the solar activity modulated by the tilt angle distribution seems to be responsible for the reversal asymmetry and its quasi-periodicity reported in this paper.

## 5. Summary and Conclusions

In this paper, we have used microwave imaging observations from the Nobeyama Radioheliograph to study the long-term variability of the Sun at low and high latitudes that are highly relevant to understand the solar dynamo mechanism. We used the microwave emission originating from large-scale features such as eruptive prominences and coronal holes. Microwave emission from active regions is also a good indicator of sunspot activity. The unique appearance of prominence eruptions at high latitudes marks the duration of the maximum phase of the solar cycle. We have used the locations of prominence eruptions as a proxy to the locations of filaments at various latitudes. The end of the high-latitude prominence activity signifies the time of sign reversal in each hemisphere. The polar reversal coincides with the polar microwave brightness enhancement above the quiet-Sun level. The brightness enhancement disappears during the maximum phase, being 180º out of phase with the sunspot cycle. The main conclusions of this study can be summarized as follows.

1. The microwave butterfly diagram and the locations of prominence eruptions obtained from microwave images provide important clues to the understanding of the long-term behavior of solar activity.

2. The microwave brightness temperature (Tb) in the polar regions is highly correlated with the polar photospheric magnetic field strength B, yielding a regression line: $B = 0.0088 Tb - 93$ G, where Tb is in K.

3. The polar microwave brightness temperature is significantly correlated to the speed of the fast solar wind. When the brightness temperature falls to the quite Sun values, only slow solar wind can be found as confirmed from Ulysses/SWOOPS observations.

4. A cross-correlation analysis shows that the polar brightness temperature in cycle n is correlated with the strength of the cycle n+1, applicable to each hemisphere separately. The typical lag is about half a solar cycle at maximum correlation.

5. The polar microwave brightness temperature can be used as a proxy to the polar magnetic field strength in predicting the strength of solar cycles. We predict the smoothed and unsmoothed sunspot numbers to be 116 and 89 in the southern hemisphere. The lower limits to the corresponding numbers in the northern hemisphere are predicted as 97 and 59.

6. The microwave brightness temperature is not sharply peaked during the minimum. The flat appearance is somewhat similar to the flat axial dipole moment reported by Iijima et al. (2017) for three cycles. Accordingly, our prediction of the cycle-25 strength agrees with theirs. The flatness means the brightness temperature well before the solar minimum can be used for solar cycle prediction.

7. The north-south asymmetry of sign reversal at solar poles inferred from the end of rush to the poles exhibits a quasi-periodicity (3-5 cycles).

8. Two of the three reversals happened when there was prolonged high-latitude prominences most likely caused by alternating streams of positive and negative polarity surges from the trailing portions of active regions with significant contribution from regions that do not obey Joy's law.

**Competing interests:**
 The authors have no competing interests to declare.

**Acknowledgments**

NoRH is currently operated by the Nagoya University in cooperation with the International Consortium for the Continued Operation of the Nobeyama Radioheliograph (ICCON). This work utilizes SOLIS (Synoptic Optical Long-term Investigations of the Sun) data obtained by the NSO Integrated Synoptic Program (NISP), managed by the National Solar Observatory, which is operated by the Association of Universities for Research in Astronomy (AURA), Inc. under a cooperative agreement with the National Science Foundation. The Ulysses data were obtained from the final archive located at http://ufa.esac.esa.int/ufa/. The aa index data were from the National Geophysical Data Center (NGDC, ftp://ftp.ngdc.noaa.gov/STP/GEOMAGNETIC_DATA/AASTAR/aaindex) and the British Geological Survey (http://www.geomag.bgs.ac.uk/data_service/data/home.html). We thank the staff of the Kodaikanal Solar Observatory in providing access to the daily drawings of prominences and filaments; in particular we thank Prabhu Ramkumar, R. Selvendran, and Ebenezer Chellasamy for their assistance in the data used in this paper.  This work benefited from NASA's open data policy in using SOHO/MDI, SDO/AIA and SDO/HMI data. Work was supported by NASA's LWS TR&T program.